\documentclass[aps,prl,reprint,showpacs,noshowkeys,floatfix]{revtex4-1}

\usepackage{balance}

\usepackage{float}
\usepackage{graphicx}

\usepackage{amsmath}
\usepackage{mathtools}
\usepackage{amssymb}
\usepackage{amsfonts}

\usepackage[backref=none]{hyperref}
\usepackage[usenames,dvipsnames]{color}
\definecolor{MyDarkBlue}{rgb}{0,0.1,0.7}
\hypersetup{pdfborder={0 0 0},colorlinks,breaklinks=true,
  urlcolor={MyDarkBlue},citecolor={MyDarkBlue},linkcolor={MyDarkBlue}
}

\def\w{\epsilon}
\def\ext{back_o0.6_8p}

\newcommand{\rmi}{\mathrm{i}}
\newcommand{\rme}{\mathrm{e}}
\newcommand{\rmd}{\mathrm{d}}

\DeclareMathOperator*{\Tr}{Tr}
\DeclareMathOperator*{\Imag}{Im}
\DeclareMathOperator*{\Real}{Re}

\newcommand{\eref}[1]{(\ref{#1})}
\newcommand{\Leref}[1]{(\ref{#1}) of the article}
\newcommand{\fref}[1]{Fig.~\ref{#1}}
\newcommand{\Lfref}[1]{Fig.~\ref{#1} of the article}

\newcommand{\Fref}[1]{Figure~\ref{#1}}

\begin{document}

\title{Quantum graphs whose spectra mimic the zeros of the Riemann zeta function}
\date{\today}

\author{Jack Kuipers}
\email{jack.kuipers@ur.de}
\author{Quirin Hummel}
\email{quirin.hummel@ur.de}
\author{Klaus Richter}
\affiliation{Institut f\"ur Theoretische Physik, Universit\"at Regensburg, D-93040
Regensburg, Germany}

\begin{abstract}
One of the most famous problems in mathematics is the Riemann hypothesis: that the non-trivial zeros of the Riemann zeta function lie on a line in the complex plane. One way to prove the hypothesis would be to identify the zeros as eigenvalues of a Hermitian operator, many of whose properties can be derived through the analogy to quantum chaos. Using this, we construct a set of quantum graphs that have the same oscillating part of the density of states as the Riemann zeros, offering an explanation of the overall minus sign. The smooth part is completely different, and hence also the spectrum, but the graphs pick out the low-lying zeros.
\end{abstract}


\maketitle

The Riemann zeta function $\zeta(s)$ encodes the distribution of the prime numbers and therefore plays a central role in number theory.
It is the analytic continuation of the infinite sum over integers $\sum_{n=1}^{\infty}n^{-s}$ to \(\Real{s} \leq 1\).
In fact its zeros can be used to obtain the prime counting function with Heaviside steps at each prime.
The famous Riemann hypothesis states that all the zeros, aside from the \emph{trivial} zeros at the negative even numbers,  lie on a \emph{critical} line with real part $1/2$.
Writing the non-trivial zeros, which come in complex conjugate pairs, as $s_{n}=1/2\pm\rmi t_{n}$, the $t_n$ would all be real if the Riemann hypothesis is true.
Proving the hypothesis remains one of the outstanding problems in mathematics, and would also prove the many propositions based on it.

Despite its number-theoretical background, the Riemann zeta function appears in the study of a range of different physical systems~\cite{sh11} including recently the freezing transition in random energy landscapes~\cite{fhk12,fk13published}.
In a connection dating to Montgomery~\cite{montgomery73} and Dyson, the zeros are particularly strongly related to the eigenvalues of random matrices from the Gaussian unitary ensemble.
Along with numerical evidence that they share the same statistics~\cite{odlyzko89}, random matrices have also been used to obtain conjectures about the moments of $\zeta(s)$~\cite{ks00}.

Already Hilbert and P\'olya recognized (as discussed in~\cite{montgomery73}) that if the $t_n$ could be identified as the eigenvalues of a Hermitian operator, they would necessarily be real and the hypothesis proved.
This has triggered the search for a quantum system described by such a Hermitian Hamiltonian.
Moreover, the fact that the spectra of quantum systems with chaotic classical counterparts obey random matrix statistics~{\cite{haake}}, like the Riemann zeros, suggests seeking an appropriate quantum chaotic system.
Further hints~{\cite{bk99}} about the nature of such a system derive from an analogy starting from the density of states.
Placing a delta function at each zero along the critical line, the density of states $d(t)=\sum_{n=1}^{\infty} \delta(t-t_n) = \bar{d}(t) + d^{\mathrm{osc}}(t)$ can be considered as consisting of two parts: a smooth average background $\bar{d}(t)$ overlaid with an oscillating part $d^{\mathrm{osc}}(t)$.
For the zeros of the Riemann zeta function the smooth part (see \textit{e.g.}~{\cite{bk99}})
\begin{equation} \label{riemanndbar}
  \bar{d}(t)=\frac{1}{2\pi}\ln\frac{t}{2\pi} + O\left(\frac{1}{t^2}\right) \,,
\end{equation}
is logarithmically increasing while the oscillating part has the divergent expression {\cite{berry85, bk99}}
\begin{figure}
	\begin{flushleft}
		\includegraphics[width=8.5cm]{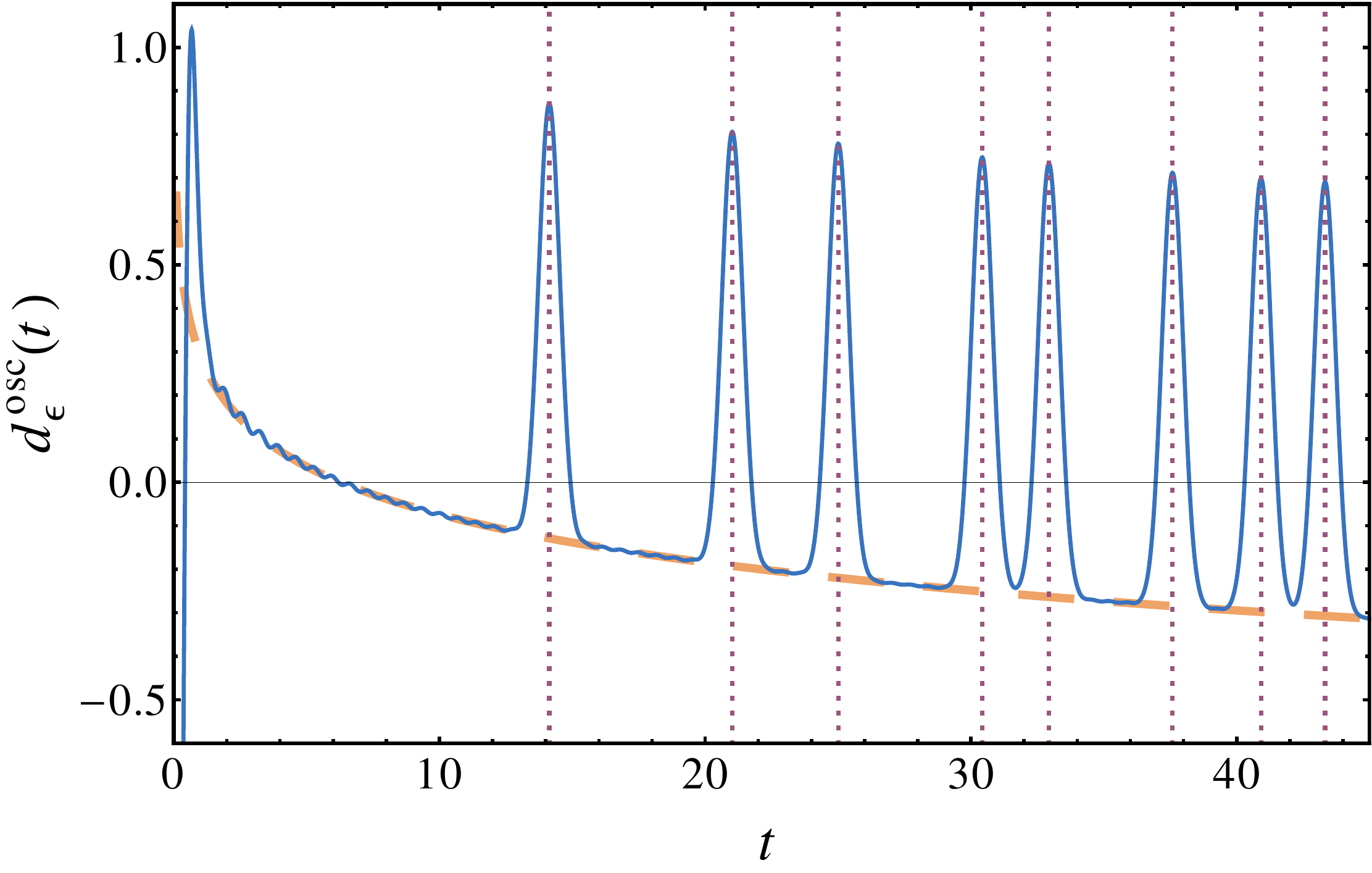}
		\caption{\label{fig:dfl_riemann}Solid blue: oscillating part \eref{riemanndosc} of the density of zeros of the Riemann zeta function smoothed with a Gaussian of width $\w=0.4$. The sums over $p$ and $m$ are truncated to $\w m \ln p \leq 3.7$.
		Dotted purple: exact zeros of the Riemann zeta function.
		Dashed orange: negative of smooth part \eref{riemanndbar}.}
	\end{flushleft}%
\end{figure}
\begin{equation} \label{riemanndosc}
  d^{\mathrm{osc}}(t)=-\frac{1}{\pi}\sum_{p}\sum_{m=1}^{\infty}\frac{\ln p}{p^{m/2}}\cos(tm\ln p) \,,
\end{equation}
involving a sum over all primes $p$ and their `repetitions' $m$.
One can enforce convergence by smoothing with a Gaussian of width $\w$, washing out terms with $\w m \ln(p) \gg 2\pi$.
Plotting the smoothed sum of the remaining primes in \fref{fig:dfl_riemann} we see peaks exactly at the Riemann zeta zeros along with the overall divergence at $t=0$.
As $t$ increases, the zeros come closer according to~\eref{riemanndbar}, so with fixed $\w$ we eventually stop being able to resolve them.  The mean part of the density of states can also be seen in \fref{fig:dfl_riemann} as the lower curve needed to map the gap between the peaks to the axis.
See the Appendix for the density in the complex plane. 

The formula in~\eref{riemanndosc} is remarkably similar to Gutzwiller's trace formula~\cite{gutzwiller71} for the oscillating part of the density of energy states of chaotic quantum systems
\begin{equation} \label{gutzwillertraceformula}
		d^{\mathrm{osc}}(E)\approx\frac{1}{\pi \hbar}\sum_{\gamma} \sum_{m=1}^{\infty} T_{\gamma} A_{\gamma,m} \cos\left(\frac{m S_{\gamma}}{\hbar} - \frac{m \pi \mu_{\gamma}}{2} \right) \,.
\end{equation}
The sum is over the classical primitive periodic orbits $\gamma$ of the system and their repetitions $m$.
Each orbit has period $T_{\gamma}$, stability amplitude $A_{\gamma}$, reduced action $S_{\gamma}$ and Maslov index $\mu_{\gamma}$.
The close correspondence between~{\eref{gutzwillertraceformula}} and~{\eref{riemanndosc}} becomes evident if we associate primes $p$ in~{\eref{riemanndosc}} with primitive periodic orbits $\gamma$ in~{\eref{gutzwillertraceformula}}.

For a quantum chaotic system to match the Riemann zeta function, a wide range of properties can be deduced from this analogy including that the dynamics should be quasi one dimensional and without time reversal symmetry~\cite{bk99}.
Berry and Keating realized that many of the properties are satisfied by the simple Hamiltonian $H=xp$, except that the motion is unbounded~\cite{bk99}.
With a truncation near the origin, the semiclassical mean density of states is also the same as \eref{riemanndbar} but the problem was in finding boundary conditions to give a Hermitian operator with real eigenvalues.
Various related operators have since been obtained which keep the same mean part of the density of states as the Riemann zeros~\cite{st08,srednicki11,srl11,bk11,sierra12}.
However these extensions still miss two of the trickiest properties that the quantum system should have and their spectra do not match the Riemann zeros.

A closer comparison of~\eref{gutzwillertraceformula} to~\eref{riemanndosc} highlights these two properties.  Namely that the periodic orbits should (i)~have primitive lengths $\ln(p)$ corresponding to all primes $p$ and (ii)~all have a Maslov phase of $\pi$ to obtain the minus sign in~\eref{riemanndosc}.  The Maslov phase is the term $m \pi \mu_{\gamma}/2$ in~\eref{gutzwillertraceformula} so the second requirement cannot hold for all $m$.  Overcoming this contradiction is the aim of this paper.

To enforce that the periodic orbits only have particular lengths of $\ln(p)$ we turn to quantum graphs where the Schr\"odinger equation acts on the bonds of a network.
Quantum graphs arise as simplified models in a range of physical applications~{\cite{bk13}}, providing for example the spectra and dispersion relations of carbon nano-structures like graphene~{\cite{kp07}}, and fittingly are used to capture the essence of the quantum behavior of chaotic systems~{\cite{ks97,gs06}}.
The bonds are one dimensional so we directly satisfy one of the properties needed while we later tune the lengths and connections between the bonds to obtain the two requirements listed above. Focusing on the oscillating part of the density of states, we then construct infinite graphs which match~\eref{riemanndosc}.  Since the zeros of the Riemann zeta function come in complex conjugate pairs with $\pm t_n$, they behave more like a wavenumber than energy spectrum [used in the analogy with~\eref{gutzwillertraceformula}].
We therefore now treat $t_n$ as wavenumbers \(k\), especially since the trace formula for quantum graphs involves $k$ directly.
This does not affect the analogy since the $k$, as square roots of positive real energies, correspond to a Hermitian operator.

\textit{Quantum graphs}.---The density of states for a quantum graph has a mean part $\bar{d}(k)=L_{\rm tot}\slash\pi$ given by the total length $L_{\rm tot}$ of the bonds, counting directed bonds with a factor of \(1/2\).  This is independent of $k$, unlike for the Riemann zeta zeros in~\eref{riemanndbar}, but the oscillating part of the density of states \cite{gs06} reduces to
\begin{equation} \label{graphdosc}
  d^{\mathrm{osc}}(k)=+\frac{1}{\pi}\sum_{\gamma}\sum_{m=1}^{\infty}L_{\gamma}A_\gamma^m \cos(km L_{\gamma}) \, ,
\end{equation}
if \(\sum_{m L_\gamma = L} L_\gamma A_\gamma^m \in \mathbb{R}\) for all possible periodic orbit lengths \(L\).
The length $L_{\gamma}=\sum_{e\in \gamma} L_{e}$ of the primitive orbit \(\gamma\) is the sum of the lengths of the edges involved and $A_\gamma^m=\prod \sigma_{e_{i+1},e_{i}}^{m}$ the product of the scattering matrix elements the orbit passes through with $e$ being the edges in $\gamma$.

One can then construct a quantum graph with periodic orbits of length $\ln(p)$ by simply setting the lengths of the bonds to the same value, but this gives two problems:
(a)~If orbits of length $\ln(p)$ and $\ln(q)$ connect at the same vertex, we can have an orbit of length $\ln(pq)$ which is a composite number.
(b)~If an orbit corresponding to a prime $p$ has a negative prefactor \(A_{\gamma}<0\), its repetitions have prefactors \(A_\gamma^m\) giving the even repetitions the wrong sign.

\begin{figure}
	\begin{flushleft}
		\includegraphics[width=8.5cm]{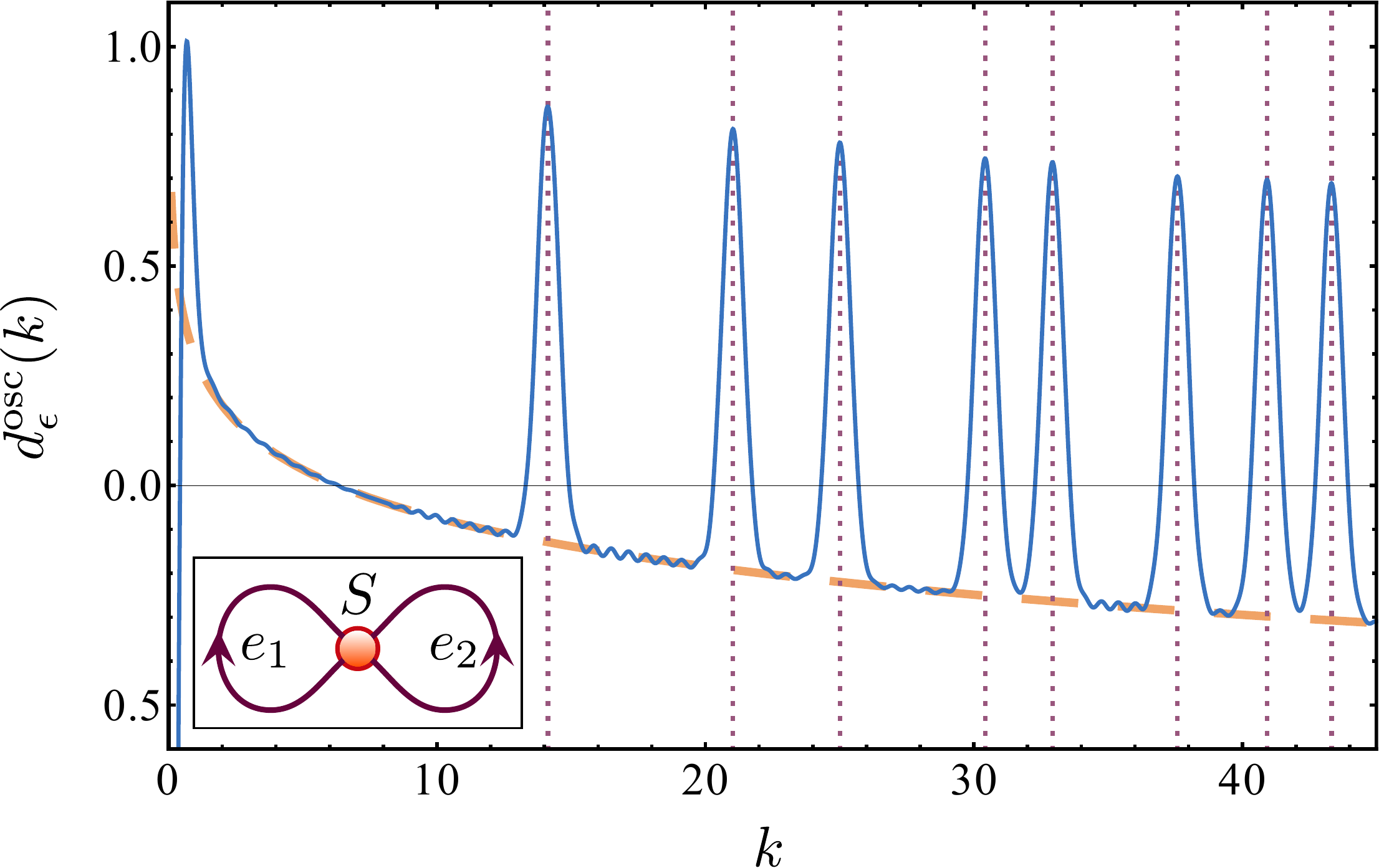}
		\caption{\label{fig:dfl_butterfly}Solid blue: density of exact eigenvalues of the set of butterfly graphs~\eref{scatteringrecursion4} smoothed with a Gaussian of width $\w=0.4$ and the mean part subtracted. The set of graphs is truncated at $\w m \ln p \leq 3.7$.
		Dotted purple: exact zeros of the Riemann zeta function.
		Dashed orange: negative of the smooth part~\eref{riemanndbar} of the Riemann zeros density.
		Inset: A butterfly graph. Two directed bonds $e_1$ and $e_2$ of the same length $L$ meet at a single vertex, characterized by a $2\times2$ scattering matrix $S$.}
	\end{flushleft}
\end{figure}

To avoid problem (a), we can simply only connect orbits whose lengths only involve the same prime $p$.  In particular we can consider \emph{butterfly} graphs made up of two identically long directed bonds which meet at a single vertex.
The inset of \fref{fig:dfl_butterfly} shows such a butterfly graph.

Along its directed bonds the wavefunction admits the solutions $\varphi_1(x_1)=c_1 {\rm e}^{\rmi k x_1}$ and $\varphi_2(x_2)=c_2 {\rm e}^{\rmi k x_2}$, where $x_i$ are the coordinates along the bonds starting at the vertex and following the direction of the bond.
At the vertex, the wavefunction has to fulfill boundary conditions
\begin{equation} \label{bndry cond}
	\left( \begin{array}{c}
					c_1 \\
					c_2
	       \end{array} \right) =
	S \left( \begin{array}{c}
					c_1 \rme^{\rmi kL}\\
					c_2 \rme^{\rmi kL}
	       \end{array} \right)
	=	\left( \begin{array}{cc}
					\sigma_{11} & \sigma_{12}\\
					\sigma_{21} & \sigma_{22}
	       \end{array} \right)
			\left( \begin{array}{c}
							c_1 \rme^{\rmi kL}\\
							c_2 \rme^{\rmi kL}
						 \end{array} \right) \,,
\end{equation}
defined by a scattering matrix $S$, where $L$ is the length of each bond.
Each element $\sigma_{ji}$ is the scattering amplitude for an incoming wave on bond $e_i$ to an outgoing wave on bond $e_j$.
In order to preserve total probability current during the scattering process, $S$ needs to be unitary which makes the Hamiltonian self-adjoint [since condition~\eref{bndry cond} also implies that the first derivatives of the \(\varphi_j\) are connected by \(S\)].

The scattering matrix can therefore be written as
\begin{equation} \label{Stotheta}
	S = U^\dagger
			\left(\begin{array}{cc}
       \rme^{\rmi \theta_1} & 0 \\
       0 & \rme^{\rmi \theta_2}
      \end{array}\right)
			U \,,
\end{equation}
where $U$ is a unitary matrix and $\rme^{\rmi \theta_j}$ are the eigenvalues of $S$.
For a given butterfly graph with scattering matrix $S$ and bond length $L$, the wavenumbers $k$ which admit wavefunctions on the graph satisfy \(\det (1-\rme^{\rmi k L}S) = 0 \).
In terms of the eigenvalues $\rme^{\rmi \theta_{j}}$ of $S$, this simply means \(k = (2\pi z - \theta_{j}) \slash L \) for all integer $z$ so we have a periodic pair of solutions, corresponding to the mean density $L\slash\pi$.
To plot the corresponding spectra of the graphs derived later we subtract the mean part and smooth them with some width $\w$.
Terms with $\w L \gg 2\pi$ can then be excluded from the calculation.

If $\Tr S$ is real we have the further simplification that $\theta_1=-\theta_2$ (or \(\theta_1=\theta_2 + \pi\) for vanishing \(\Tr S\)) and a usefully simple form of the trace formula for butterfly graphs
\begin{equation} \label{butterflygraphdosc}
  d^{\mathrm{osc}}(k)=+\frac{1}{\pi}\sum_{m=1}^{\infty}L \Tr S^m \cos(km L) \, .
\end{equation}

\textit{Quantum graphs and the sign problem}.---The bigger problem is the sign problem of creating orbits with the correct Maslov phase.  To overcome this problem we can view the contribution of each prime $p$ as coming from an (infinite) family of graphs that together give the required phase rather than being the result of a single orbit.  We now turn to constructing such a set of graphs.

We start by trying to obtain the contribution to the trace formula~\eref{riemanndosc} for a single prime, say $p=2$ and we aim to build a graph that mimics the corresponding term:
\begin{equation} \label{pis2term}
-\frac{1}{\pi}\sum_{m=1}^{\infty}\frac{\ln 2}{\sqrt{2}^{m}}\cos(km\ln 2) \, .
\end{equation}
Comparing the $m=1$ term to that of~\eref{butterflygraphdosc}, a butterfly graph with length $L=\ln(2)$ and scattering matrix with $\Tr S = -1/\sqrt{2}$ would directly give agreement.  Labeling the scattering matrix by the subscripts $(p,m)$ for consistency later, we can simply set it to be
\begin{equation} \label{unitaryprime2rep1}
S_{2,1}=\left(\begin{array}{cc}
       -\frac{1}{2\sqrt{2}} & \frac{\sqrt{7}}{2\sqrt{2}} \\
       -\frac{\sqrt{7}}{2\sqrt{2}} & -\frac{1}{2\sqrt{2}}
      \end{array}
\right) \,,
\end{equation}
though a more general form exists.  This butterfly graph then gives a contribution involving $\Tr S_{2,1}^2=-3/2$ when $m=2$ in~\eref{butterflygraphdosc}.  This is composed of $1/8$ from each second repetition of the bonds of length $\ln(2)$ and $-7/4$ from the orbit that covers them both (with twice the primitive length).  However, \eref{pis2term} suggests we want a contribution of $-1/2$ instead.  We can then add a second uncoupled butterfly graph with its own trace formula~\eref{butterflygraphdosc} with two bonds of length $L=\ln(4)$ and a scattering matrix of
\begin{equation} \label{unitaryprime2rep2}
S_{2,2}=\left(\begin{array}{cc}
       \frac{1}{4} & \frac{\sqrt{15}}{4} \\
       -\frac{\sqrt{15}}{4} & \frac{1}{4}
      \end{array}
\right) \, ,
\end{equation}
or any unitary matrix with $\Tr S_{2,2} = 1/2$.  As $\ln(4)=2\ln(2)$, this effectively gives an additional $+1$ to the $-3/2$ contribution from the second power of \eref{unitaryprime2rep1} and therefore the required value.

For $m=3$ and the orbits of length $\ln(8)$, we obtain a contribution of $5\sqrt{2}/4$ from $\Tr S_{2,1}^3$ while we would want $-\sqrt{2}/4$.  If we add a further uncoupled butterfly graph with two bonds of length $\ln(8)$ and with the same vertex scattering matrix as in~\eref{unitaryprime2rep1} they would add an additional $-3/\sqrt{2}$ to give the required contribution of $m=3$ in \eref{pis2term}.

It looks like we could continue this process and define for each prime $p$ a set of independent butterfly graphs $(p,m)$ each with two bonds of equal length $m \ln(p)$ for all $m \in \mathbb{N}_+$.  Each pair of bonds is coupled by unitary $2\times2$ scattering matrices $S_{p,m}$ with diagonal entries defined so that the whole set conspires to give the sum in~\eref{pis2term}.

The relations that the butterfly graphs need to satisfy are
\begin{equation} \label{scatteringrecursion}
\sum_{d \vert m} d\Tr S_{p,d}^{m/d} = -\frac{1}{p^{m/2}} \,,
\end{equation}
involving a sum over divisors of $m$ so that each matrix could be obtained recursively:
\begin{equation} \label{scatteringrecursion2}
	\Tr S_{p,m} = -\frac{1}{m p^{m/2}} - \sum_{d \vert m}^{d < m} \frac{d}{m} \Tr S_{p,d}^{m/d}  \,.
\end{equation}
A problem arises however because the sum of divisors of $m$ (divided by $m$) is unbounded since
\begin{equation}
	\limsup_{m \rightarrow \infty}  \frac{\sum_{d \vert m} \frac{d}{m}}{\log \log m} = \rme^{\gamma_{\rme}} \,,
\end{equation}
where $\gamma_{\rme}$ is the Euler-Mascheroni constant.
Thus the trace of $S_{p,m}$ might need to be larger than 2 in absolute value. For example this happens for $m=24$ for the primes $3$ and $5$ (among others) allowing no solution of~\eref{scatteringrecursion2}. However, we may simply add additional identical copies of the graphs with that length to share the trace between them and find a solution.

Since the traces of all $S_{p,m}$ need to be real [as indicated by \eref{scatteringrecursion2}], we can replace $S_{p,m}$ by some unitary matrices $U$ and real angles $\pm\theta_{p,m}$ as in~\eref{Stotheta} (in case of vanishing \(\Tr S\) we are free to choose \(\theta=\pm \pi / 2\) among others).  The spectra of the graphs only depend on the $\theta_{p,m}$ and not on the specific choice of the $U$ so we work with the angles directly.
The following prescription then defines the eigenvalues of the whole graph including $l_{p,m}$ copies of the butterfly graph $(p,m)$ recursively:
\begin{equation} \label{scatteringrecursion4}
		T_{p,m} \equiv \frac{1}{2m p^{m/2}} + \sum_{d \vert m}^{d < m} \frac{d}{m} l_{p,d} \cos\Big(\frac{m}{d} \theta_{p,d}\Big) \,,
\end{equation}
where $\cos(\theta_{p,m}) = T_{p,m} \slash l_{p,m}$ , $l_{p,m} = \big\lceil \vert T_{p,m} \vert \big\rceil$, and \(\lceil \cdot \rceil\) denotes the ceiling function.
Numerically, in the range of $m$ and $p$ we explored, the occurrence of pairs $(p,m)$ requiring $l$ copies decays strongly with $l$.
A value of $l = 3$ is already very rare [the first occurrence was at $(p,m) = (3,1710)$].
With a fixed cutoff $m<M$ and $p<P$ it is of course always possible to find solutions of~\eref{scatteringrecursion4} in that range.
This prescription then gives an identical oscillating part of the density of states as~\eref{riemanndosc} up to the cutoff, and differences thereafter.  \Fref{fig:dfl_butterfly} shows the fluctuating part of the spectrum of a truncated set of butterfly graphs constructed in this manner which can be compared to \fref{fig:dfl_riemann} involving the primes. The difference in the damped high frequency oscillations is due to the truncation and would vanish when sending the cutoff to infinity. None of the partial graphs $(p,m)$ used for \fref{fig:dfl_butterfly} needed copies to be taken into account.

Although in principle the $l_{p,m}$ could be unbounded, this only adds additional graphs to an already infinitely large set. Nevertheless, to obtain better control over the numbers of copies we could also construct a different set of butterfly graphs by setting $l_{p,m}=m$.
An alternative approach only uses bonds of length $\ln(p)$ for each $p$.
They are detailed in the Appendix.

\textit{Quantum graphs and the Riemann zeros}.---We then have several constructions for each prime whereby infinite sets of graphs (with two bonds each) together match the oscillating contribution to the density of states that the prime contributes for the zeros of the Riemann zeta function.  Combining the sets for all primes then leads to a swarm of butterfly graphs which (like the primes) pick out the Riemann zeros.

The constructions offer an explanation for the puzzling properties that a Riemann quantum chaotic system should possess -- namely that the orbits should have lengths $\ln(p)$ for primes $p$ and their repetitions should all have a Maslov phase of $\pi$.  The quantum graphs show that the Maslov phase of $\pi$ can actually derive from different orbits of the same length working together.

The possibility of the phase deriving from many orbits has previously been hypothesized in~\cite{srednicki11prl}. There Andreev reflection automatically provides the dominant periodic orbits and their odd repetitions with a Maslov phase of $\pi$ while the even repetitions would have a phase of $0$ and the opposite sign.  However, it was noted that including orbits of length $2^l \ln(p)$ for all $l \in \mathbb{N}$ could in principle compensate for the even repetitions. An alternative explanation for the overall minus sign in~\eref{riemanndosc} is that the zeros correspond to an absorption spectrum~\cite{connes99} although this removes the necessity for the $t_n$ to be real.

Each butterfly graph has a simple periodic pair of wavenumbers unconnected to the Riemann zeros.
Combining a set of uncoupled butterfly graphs provides a composite object whose spectra is a superposition of the individual periodic spectra and whose wavefunctions are localized on the corresponding butterfly.
The mean part of the swarm diverges (although it can be subtracted in a controlled way) and the spectra of their wavenumbers become infinitely dense.  Nonetheless correlations between these spectra mean the butterflies conspire to beat their wings together constructively at the Riemann zeros.

Along with the divergence of the mean density of states, a further problem is simply that the Weyl asymptotics for quantum graphs $\bar{d}(k)=L_\mathrm{tot}\slash\pi$ is independent of $k$, unlike for the Riemann zeta zeros in~\eref{riemanndbar}. The $H=xp$ operator, and its `square', have also been considered on finite quantum graphs~\cite{es10}, with the result that the mean part of the density of states likewise cannot possibly match~\eref{riemanndbar}.  Intriguingly however, infinite quantum graphs can be constructed with a logarithmic $k$-dependence like~\eref{riemanndbar} by adding bonds of decreasing length~\cite{es11}.

Here, for simplicity of the constructions, we only focused on separated graphs with bond lengths given by the primes $\ln(p)$.
The fact that we constructed several different systems which match~\eref{riemanndosc} suggests that there are many more ways to achieve this.  For example, connected orbits of other lengths could also exist as long as their contribution cancels in the end.  In particular to avoid the infinite families of graphs and to reduce the divergence in the total bond length, one could imagine connecting the graphs and reusing the bonds so they contribute to many different periodic orbits.

A simple starting point could be to connect directed bonds of
length $\ln\left(\frac{n+1}{n}\right)$ for $n\in\mathbb{N}_+$ at a single vertex
(essentially a star graph).  This automatically gives infinitely many
orbits of each length $\ln(p)$ by following all the bonds between $n$
and $np-1$; along with many other lengths which of course would need to
cancel.  Such bond lengths are similar to those which arise from
considering the full smooth part of the Riemann zeros~\cite{egger11}.
With an identity scattering matrix at the vertex, the system would also
be almost identical to the uncoupled one in~\cite{es11} rescaled to
provide the leading term in~\eref{riemanndbar}.

Returning to the butterfly graphs considered here, without the same mean part, the resulting composite spectra cannot prove the reality of $t_n$, but instead we propose that targeting the oscillating part of the density of states while trying to reduce the mean part could offer a possible route to finding a quantum system that exactly mimics the Riemann zeros.

\textit{Acknowledgments}.---This work is partly supported by the DFG (within FOR 1483). We thank Gregory Berkolaiko, Sebastian Egger and Jonathan Keating for useful discussions and critical comments.

\bibliography{qgmrz_arXiv_final}
{\color{white}
\begin{widetext}
	\vspace*{50mm}
\end{widetext}
dummy
}

\cleardoublepage

\newcommand{\sectionQ}[1]{\section{}\vspace*{-10mm}\begin{center}{\bf #1}\end{center}}
\setcounter{section}{1}

\appendix

\sectionQ{Appendix A: The density of Riemann zeta zeros in the complex plane} \label{app:complex}

Taking the logarithmic derivative \( \zeta^\prime(z) \slash \zeta(z) \) of the Riemann zeta function transforms its zeros into single poles in the complex plane.
Therefore its real part evaluated at the critical line, \( z = 1 / 2 + \rmi t - \tau \) in the limit \( \tau  \rightarrow 0^- \), gives Dirac-delta peaks at all the non-trivial zeros, assuming they indeed lie on the critical line \( z = 1/2 + \rmi t \).
The divergent expression~\Leref{riemanndosc} for the oscillating part of the density of the Riemann zeros corresponds to \( \Real [(\zeta' / \zeta)(1/2 + \rmi t)] / \pi \) using the Euler product representation of \(\zeta\) and ignoring the fact that this only converges for \( \Real z > 1 \).
Smoothing expression~\Leref{riemanndosc} with a Gaussian finally makes it convergent and transforms the delta peaks into Gaussian peaks of finite width.

In order to visualize the density of Riemann zeros in the complex plane we define
\begin{equation} \label{dosccomplex}
	d_\mathbb{C}^\mathrm{osc} (t + \rmi \tau) = - \frac{1}{\pi} \sum_p \sum_{m=1}^\infty \frac{\ln p}{p^{m/2}}
		\exp \left( - \rmi t m \ln p + \tau m \ln p \right)
\end{equation}
on the half plane \(\tau \leq 0\) to the right of the critical line, which equals \( (\zeta' / \zeta)(1/2 + \rmi t - \tau) \slash \pi \) using the divergent series expansion of \(\zeta(z)\).

To achieve convergence for \( \tau \in [-1/2, 0] \), we smooth it parallel to the critical line by convolving it with a Gaussian with respect to \(t\).
For the left side \(\tau > 0 \) of the critical line we apply the reflection property of the \( \zeta \)-function
\begin{equation}
	\zeta(1-z) = 2 (2 \pi)^{-z} \cos \left(\frac{\pi z}{2}\right) \Gamma(z) \zeta(z)
\end{equation}
on \(\zeta^\prime / \zeta = ( \ln \zeta )' \) to define
\begin{equation}
	d_\mathbb{C}^\mathrm{osc} (t+ \rmi \tau) = - [d_\mathbb{C}^\mathrm{osc} (t - \rmi \tau)]^* - 2 \bar{d}_\mathbb{C}(t + \rmi \tau) \,,
\end{equation}
which relates it back to the density evaluated at \(\tau < 0 \) using~\eref{dosccomplex}.
\( \bar{d}_\mathbb{C} \) is defined as
\begin{align}
	\bar{d}_\mathbb{C}(t + \rmi \tau) = &- \frac{1}{2 \pi} \ln 2 \pi - \frac{1}{4} \cot\left(\frac{\pi}{4} + \frac{\pi}{2}( \rmi t - \tau) \right) \nonumber\\
	&+ \frac{1}{2 \pi} \psi\left(\frac{1}{2} - \rmi t + \tau\right) \,,
\end{align}
where \(\psi(z)\) denotes the digamma function [for \( \tau = 0 \), it corresponds to the smooth part~\Leref{riemanndbar}].

\begin{figure}
		\includegraphics[width=8.5cm]{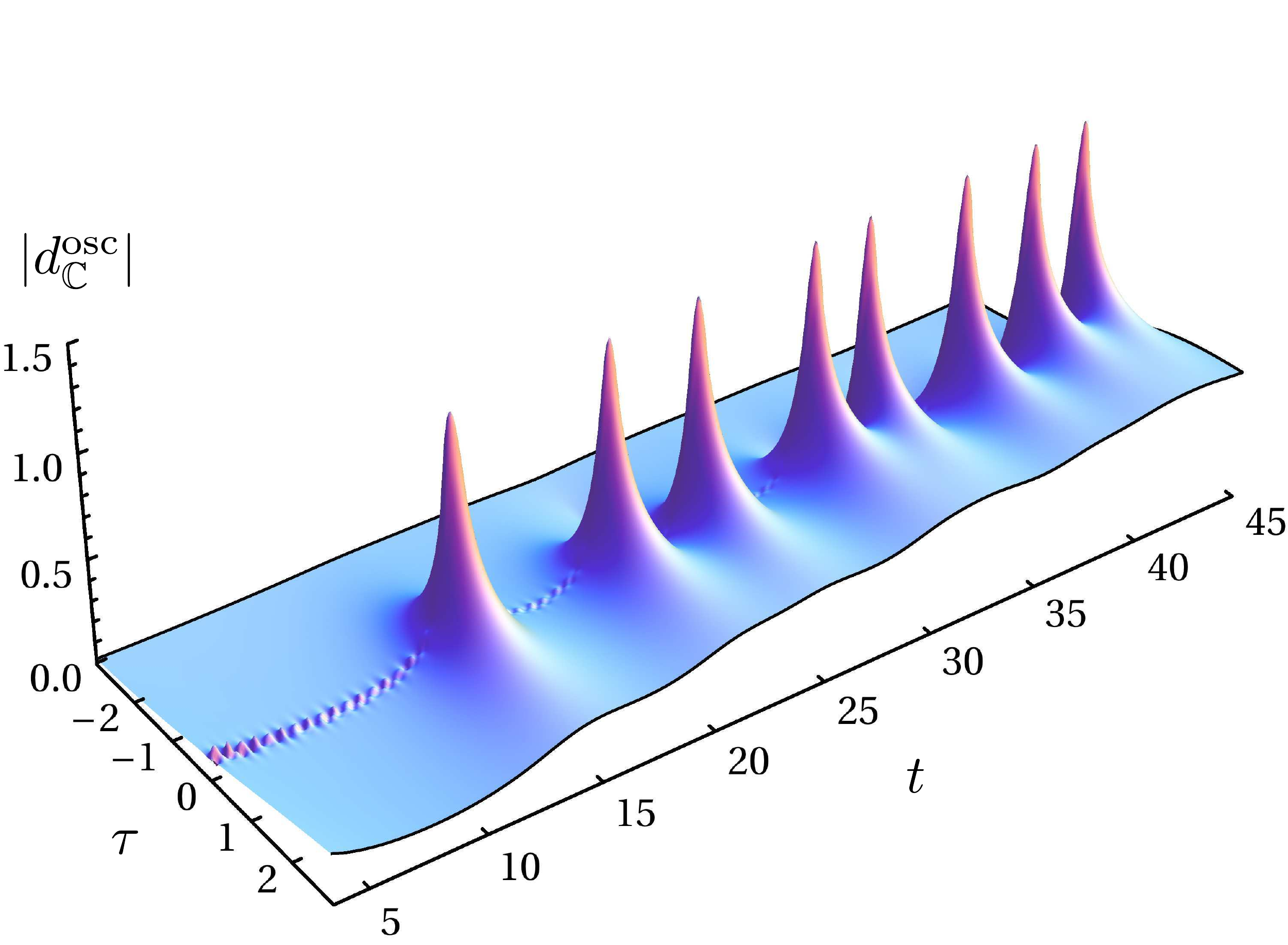}
		\caption{\label{fig:3dplotAbs} Absolute value \( |d_\mathbb{C}^\mathrm{osc} (t + \rmi \tau)| \). The first 10000 primes have been used and a smoothing in \(t\)-direction with a Gaussian of standard deviation 0.3 has been applied.}
\end{figure}

\begin{figure}
		\includegraphics[width=8.5cm]{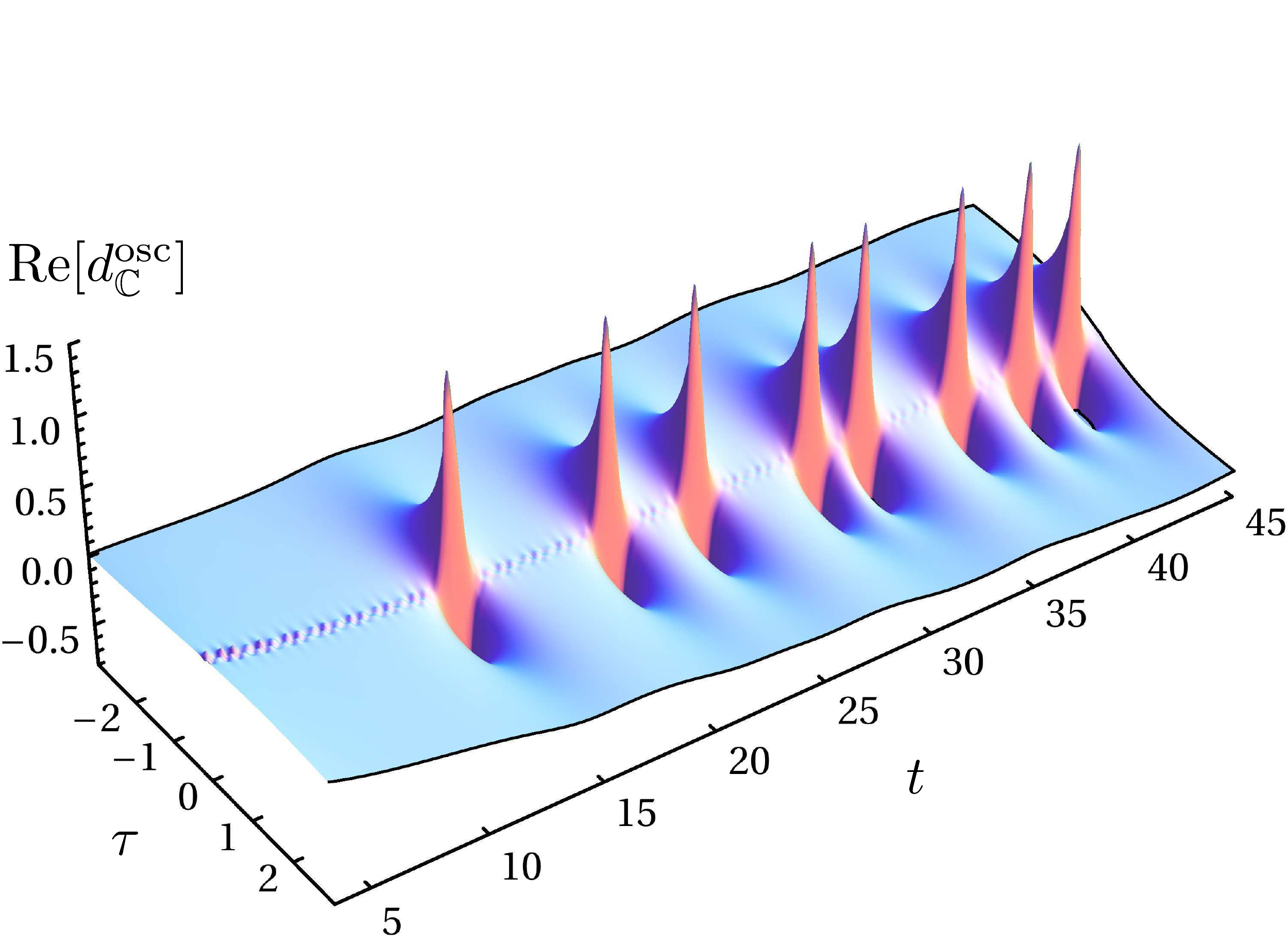}
		\caption{\label{fig:3dplotRe} Real part \( \Real [d_\mathbb{C}^\mathrm{osc} (t + \rmi \tau)] \). The first 10000 primes have been used and a smoothing in \(t\)-direction with a Gaussian of standard deviation 0.3 has been applied.}
\end{figure}

\Fref{fig:3dplotAbs} shows the absolute value of \( d_\mathbb{C}^\mathrm{osc} (t + \rmi \tau) \) with a truncation to 10000 primes and a smoothing width of 0.3, whereas \fref{fig:3dplotRe} shows the real part of the same object.
The pole-like peaks at the non-trivial Riemann zeta zeros can easily be identified.

\vfill\eject

\sectionQ{Appendix B: Increasing the number of copies of butterfly graphs} \label{app:mcopies}

\begin{figure}

		\includegraphics[width=8.5cm]{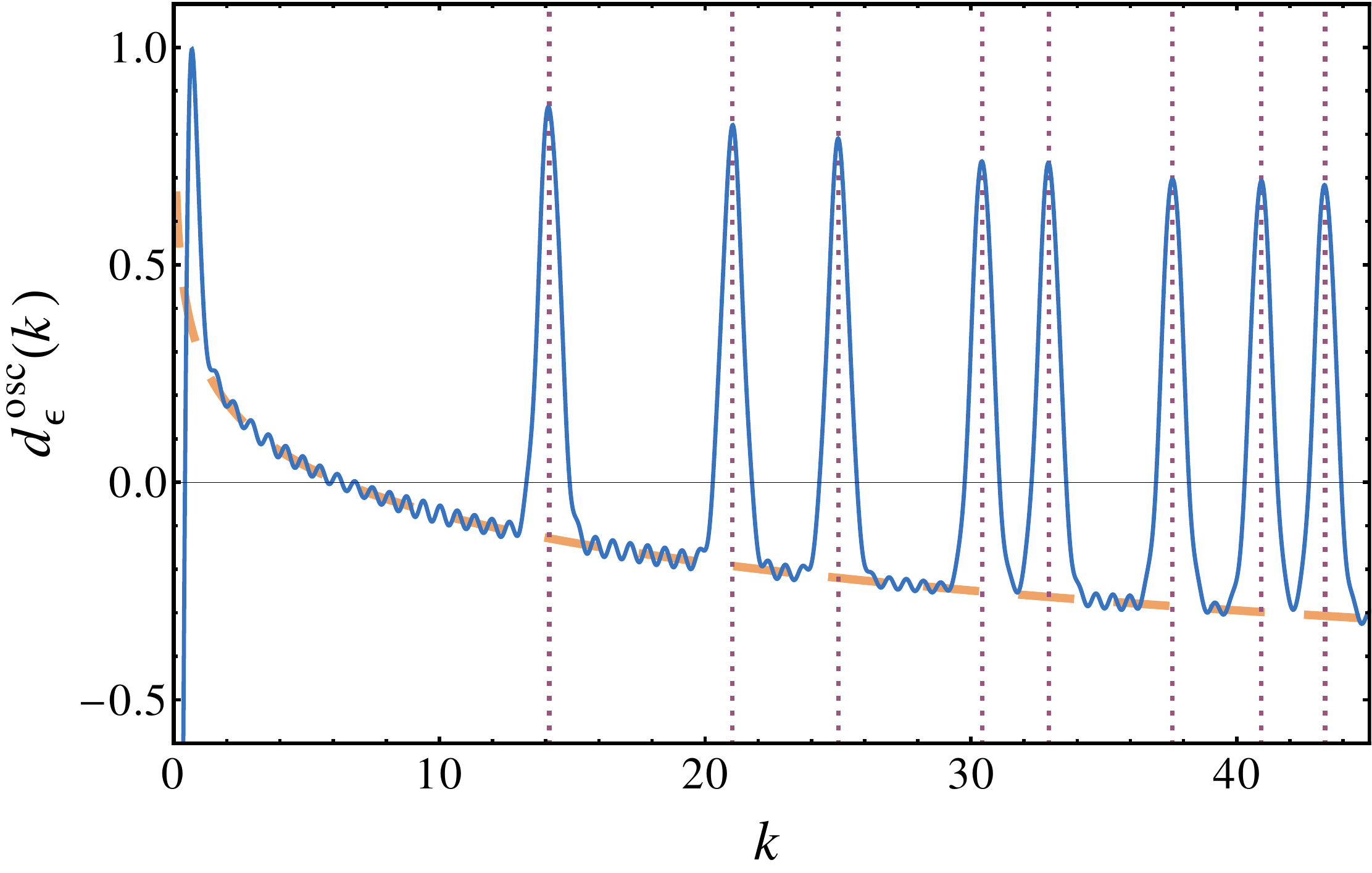}
		\caption{\label{fig:dfl_butterfly_growing}Solid blue: density of exact eigenvalues of the linearly growing set of butterfly graphs~\eref{scatteringrecursion5} smoothed with a Gaussian of width $\w=0.4$ and the mean part subtracted.
		The set of graphs is truncated at $\w m \ln p \leq 3.7$.
		Dotted purple: exact zeros of the Riemann zeta function.
		Dashed orange: negative of the mean part of the Riemann zeros density.}
\end{figure}

If from the start we set the number of copies of the butterfly graphs
to $l_{p,m}=m$ then the corresponding recursion
\begin{equation} \label{scatteringrecursion5}
	\cos(\theta_{p,m}) = -\frac{1}{2 m^2 p^{m/2}} - \sum_{d \vert m}^{d < m} \Big( \frac{d}{m} \Big)^2 \cos\Big(\frac{m}{d} \theta_{p,d}\Big) 
\end{equation}
is solvable with real $\theta_{p,m}$ for all $(p,m)$ since the sum of squared divisors of $m$ (divided by $m^2$) is bounded sufficiently.  One can estimate
\begin{eqnarray}
	\sum_{d \vert m}^{d < m} \frac{d^2}{m^2} &\leq& \sum_{d \vert m}^{\lfloor \sqrt{m} \rfloor} \frac{d^2}{m^2} + \sum_{d \vert m}^{\lfloor \sqrt{m} \rfloor} \frac{1}{d^2} - 1 \\*
	&<& \frac{1}{3\sqrt{m}} + \frac{1}{2m}+\frac{1}{6 m^{3 / 2}} + \frac{\pi^2}{6} - 1
	\equiv B(m) \,. \nonumber
\end{eqnarray}
For $m \geq 4$ the bound $B(m)$ is already smaller than 1 [$B(4) = 0.957 \ldots$] and the additional summand in~\eref{scatteringrecursion5} is $1 \slash (2 m^2 p^{m / 2}) \leq 1 \slash 128 $. Thus for all $m \geq 4$ the RHS of~\eref{scatteringrecursion5} is smaller than $1$ in absolute value. The cases $m=1,2,3$ can be easily checked by direct calculation of the sum of squares of divisors. They give sufficient bounds for~\eref{scatteringrecursion5} for the worst case \(p=2\) and hence for all primes $p$.  Therefore, the recursion~\eref{scatteringrecursion5} has solutions for all $(p,m)$.
\Fref{fig:dfl_butterfly_growing} shows the fluctuating part of the exact spectrum of a truncated set of butterfly graphs with linearly growing number of copies $l_{p,m}=m$.  The enhancement of the small damped high frequency oscillations in comparison to the butterfly graphs without linear growth in \Lfref{fig:dfl_butterfly} is a direct consequence of the increased weight of butterflies with longer bonds.

\sectionQ{Appendix C: Sets of butterfly graphs with identical lengths} \label{app:identlength}

The prescriptions~\Leref{scatteringrecursion4} and~\eref{scatteringrecursion5}
allow us to construct a sets of graphs with the same oscillating part of the density of states as the Riemann zeros, but rely on a recursive construction to find the graphs.  If we perform the sum over $m$ in Eq.~\Leref{riemanndosc} to get
\begin{equation} \label{riemanndoscsummed}
  d^{\mathrm{osc}}(t)=-\frac{1}{\pi}\sum_{p}\ln p \frac{\sqrt{p}\cos(t\ln p)-1}{p+1-2\sqrt{p}\cos(t\ln p)} \,,
\end{equation}
the resultant terms
\begin{equation}
  d^{\mathrm{osc}}_p(k)=-\frac{\ln p}{\pi} \frac{\sqrt{p}\cos(k\ln p)-1}{p+1-2\sqrt{p}\cos(k\ln p)}
\end{equation}
are periodic with period $2\pi/\ln(p)$.  This is the same period as the solutions with $m=1$ before.
A butterfly graph of bond length $\ln(p)$ contributes to $d^{\rm osc}(k)$ with this period (and also all fractions of it; the amplitudes of all the harmonics are determined by the scattering matrix).  Thus a set of butterfly graphs, all with bond length $\ln(p)$ could be constructed in a way that together they produce the correct Fourier coefficients of \(d^{\mathrm{osc}}_p(k)\).
We can try to find this set of butterfly graphs for each prime $p$ so that together they again give Eq.~\Leref{riemanndosc}.
If we label the scattering matrices of all the butterflies $S_{p,r}$, they have to fulfill the equations

\begin{equation} \label{scatteringequations}
\sum_r \Tr S_{p,r}^m = -\frac{1}{p^{m/2}} \, ,
\end{equation}
or
\begin{equation} \label{thetaequations}
2\sum_r \cos(m\theta_{p,r}) = -\frac{1}{p^{m/2}} \, ,
\end{equation}
in terms of the positive $\theta_{p,r}<\pi$.

%
\begin{figure}[t]
	\includegraphics[width=8.5cm]{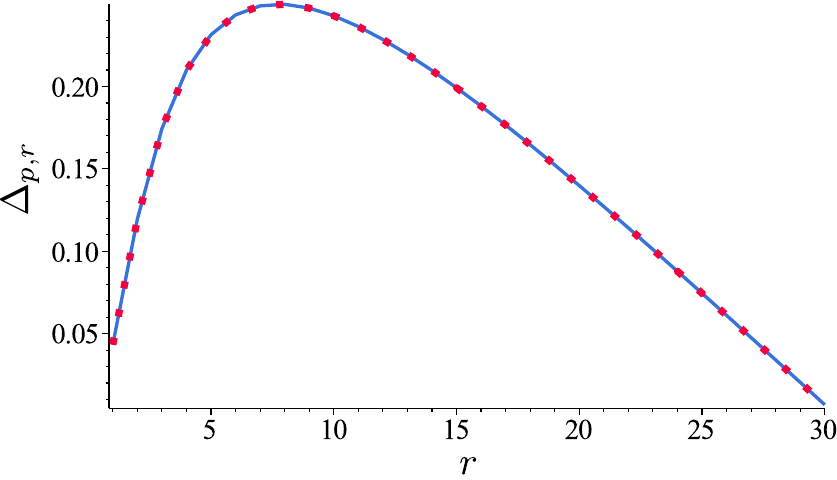}
	\caption{\label{differences30}Setting $p=2$ and $R=30$ we calculate the shift $\Delta_{2,r}$ from equally spaced angles as defined in~\eref{shifteqn}.  The dotted line is from the exact numerical solution of the 30 simultaneous equations in~\eref{thetaequations} while the solid line is derived from the approximate solutions of~\eref{approxthetas}.}
\end{figure}

\textit{Approximate Solution}.---With a finite set of graphs $r=1,\ldots,R$ we can find numerical solutions for the first $R$ equations for $m$ in~\eref{thetaequations}, but more importantly we can find approximate solutions for any number of graphs.  Assume our discrete pairs of solutions $-\pi<\pm \theta_{p,r}<\pi$ follow a density $\psi_p(\theta)$ then the limit of infinite solutions would be
\begin{equation} \label{thetadist}
2 \int_0^\pi \cos(m\theta)\psi_p(\theta) \rmd \theta = -\frac{1}{p^{m/2}} 
\end{equation}
for each $m$.  Solving for $\psi_p$ is just taking the Fourier series so
\begin{equation} \label{psiresult}
\pi \psi_p(\theta) = -\sum_m \frac{\cos(m\theta)}{p^{m/2}} = \frac{1-\sqrt{p}\cos(\theta)}{p+1-2\sqrt{p}\cos(\theta)} \, ,
\end{equation}
exactly mimicking~\eref{riemanndoscsummed}.

\begin{figure}[t]
	\includegraphics[width=8.5cm]{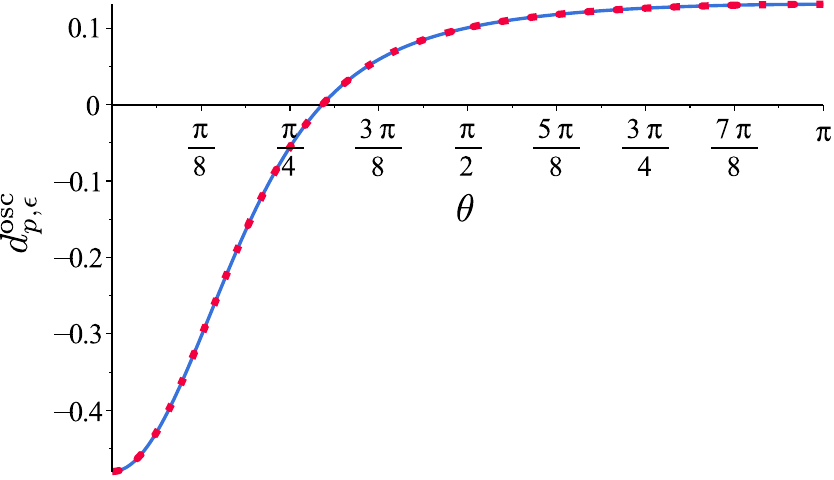}
	\caption{\label{smoothedpis2}The solid line is the density of states of 100 butterfly graphs of length $\ln(2)$ smoothed by a Gaussian function of width $w=\pi/10$ and with the mean part $100/\pi$ subtracted.  The dotted line instead is the $p=2$ term of Eq.~\Leref{riemanndoscsummed}, transformed to $\theta=k\ln(2)$ and smoothed with the same Gaussian.}
\end{figure}%

With $R$ pairs of $\theta$ solutions, the overall density should be $\psi_p(\theta)+R/\pi$ and we want discrete solutions that best approximate this density.  Dividing the overall density between \(\theta=-\pi\) and \(\theta=\pi\) into $2R$ bars of equal (unit) area, we could expect to find a solution inside each bar and we approximate by placing it at the center of mass.
Equivalently we are dividing the cumulative function evenly.  From the symmetry, we only need to look at the solutions between 0 and $\pi$ which should then satisfy
\begin{equation}
r-\frac{1}{2}=\frac{R\theta_{p,r}}{\pi}+\int_{0}^{\theta_{p,r}}\psi_p(\theta)\rmd \theta \, ,
\end{equation}
with
\begin{eqnarray} \label{integralpsi}
	\pi \int_{0}^{\theta_{p,r}}\psi_p(\theta)\rmd \theta &=& - \Imag{\sum_{m=1}^\infty \frac{\rme^{\rmi m \theta{p,r}}}{m p^{m \slash 2}}} \nonumber\\*
	&=& \Imag{ \left[ \ln\left( 1 - p^{-1\slash2} \rme^{\rmi \theta_{p,r}} \right) \right] } \,.
\end{eqnarray}
Since the argument of the logarithm in~\eref{integralpsi} always lies in the right half plane of $\mathbb{C}$, we can write
\begin{equation} \label{approxthetas}
	\pi r- \frac{\pi}{2} = R\theta_{p,r} + \arctan \frac{\sin \theta_{p,r}}{\cos \theta_{p,r} - \sqrt{p}} \,,
\end{equation}
\begin{figure}[H]
	\begin{center}
		\includegraphics[width=8.5cm]{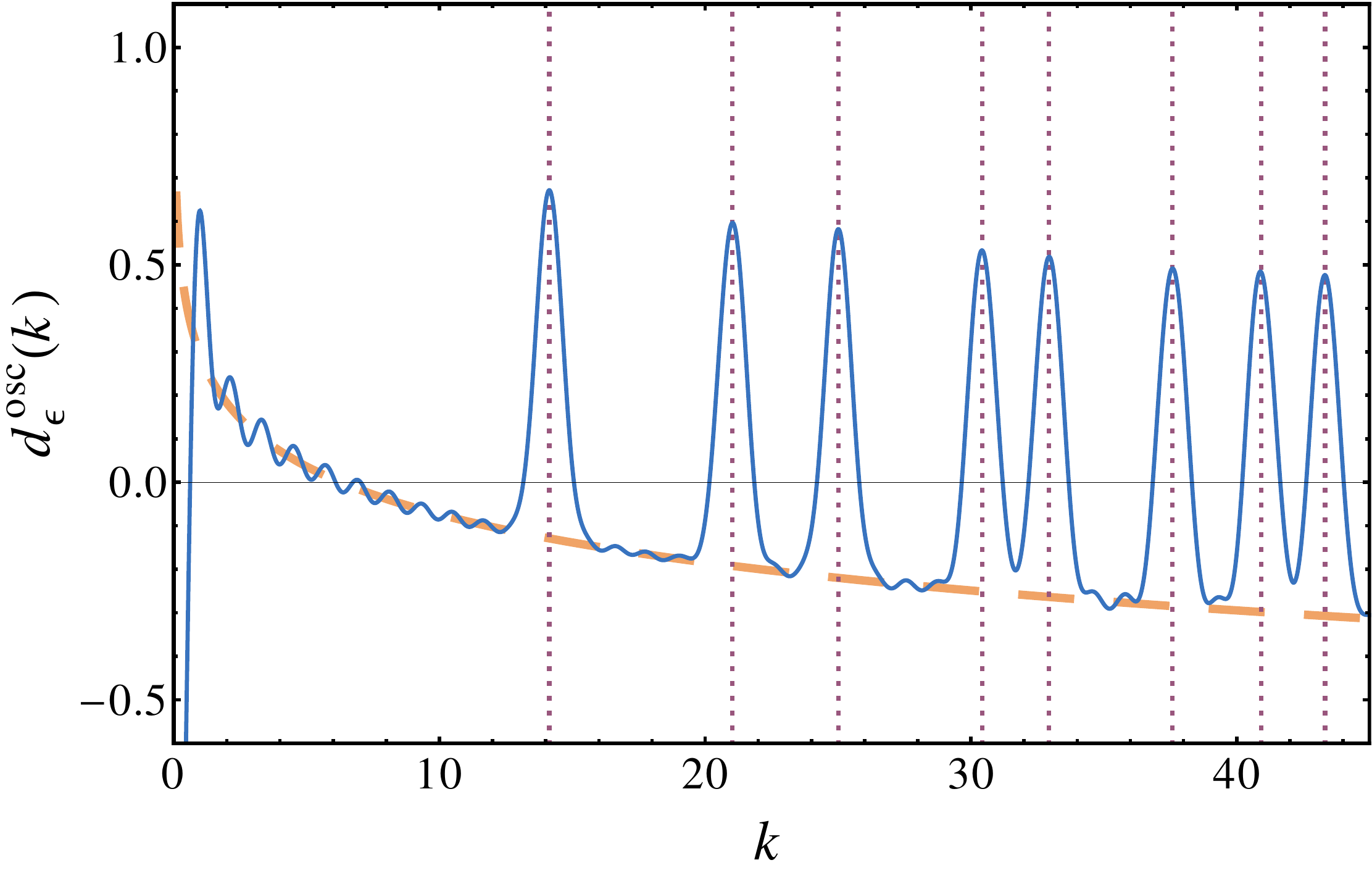}
		\caption{\label{fig:dfl_butterfly_equal}Solid blue: density of eigenvalues of the set of butterfly graphs involving identical bond lengths for each prime up to \(p=181\) derived using~\eref{approxthetas2} and the low value of $R=10$.  The result is smoothed with a Gaussian of width $\w=0.5$ and the mean part has been subtracted.
		Dotted purple: exact zeros of the Riemann zeta function.
		Dashed orange: negative of the smooth part of the Riemann zeros density.}
	\end{center}
\end{figure}%
\noindent~or equivalently
\begin{equation}  \label{approxthetas2}
	\tan (\pi \Delta_{p,r}) = \frac{\sin \theta_{p,r}}{\sqrt{p} - \cos \theta_{p,r}} \,,
\end{equation}
with \( -1 \slash 2 < \Delta_{p,r} < 1 \slash 2 \), where we defined the shift from equally spaced angles as
\begin{equation} \label{shifteqn}
	\Delta_{p,r} = \frac{R\theta_{p,r}}{\pi} - r + \frac{1}{2} \, .
\end{equation}

%
\textit{Examples}.---For $R=30$ we plot the shifts~\eref{shifteqn} for $p=2$ for both the exact solutions of the 30 equations in~\eref{thetaequations} as well as the approximate solutions from~\eref{approxthetas} or~\eref{approxthetas2} in \fref{differences30}.  At the level of the accuracy visible in the graph, these shifts are essentially identical.

Taking a larger number of approximate solutions by setting $R=100$ we place a Gaussian smoothed delta function (of width $\w=\pi/10$) on each of the $\pm \theta_{2,r}$ and their periodic repetitions and subtract the mean part $100/\pi$.  Plotting the result as the solid line in \fref{smoothedpis2} we can compare to the (equally smoothed) $p=2$ term of~\eref{riemanndoscsummed}.  In terms of angles $\theta=k\ln(2)$ this is just the convolution of $\psi_2(\theta)$ with the same Gaussian and we overlay this result as a dotted line in \fref{smoothedpis2}.  Again the lines are indistinguishable in the graph.

Even a small set of graphs is enough to be able to pick out the zeros of the Riemann zeta function.  Setting $R=10$ for example, we plot in \fref{fig:dfl_butterfly_equal} the smoothed oscillating part $d_\w^{\rm osc}(k)$ of a complete set of butterfly graphs with equal bond lengths $\ln(p)$ for primes up to 181.  The spectrum corresponds to the approximate solutions using~\eref{approxthetas2}.
\vfill

\end{document}